\documentclass[aps,pre,twocolumn,groupedaddress,draft,showpacs,showkeys]{revtex4}

\begin{document}


\title{Microscopic heat from the energetics of stochastic phenomena}


\author{Ken Sekimoto}
\email[]{ken.sekimoto@espci.fr}
\affiliation{Mati\`{e}res et Syst\`{e}mes Complexes, CNRS-UMR7057, Universit\'e
  Paris 7, France
}
\affiliation{{Gulliver,} CNRS-UMR7983, ESPCI, 
Paris, France}

\date{\today}

\begin{abstract}
The energetics of the stochastic process has shown the balance of energy on the mesoscopic level.
The heat and the energy defined there are, however, generally different from their macroscopic counterpart.
We show that this discrepancy can be removed by adding to these quantities the
reversible heat associated with the mesoscopic free energy.
\end{abstract}


\maketitle

\section{Introduction\label{sec:1}}
This Communication demonstrates the relations between 
the `mesoscopic heat' which came out in the context of the energetics of
a single realization of stochastic processes \cite{sekimoto97} (the stochastic energetics, for short)
and the (usual) heat that can in principle be measured experimentally by
calorimetric techniques or be calculated from microscopic theories or
simulations.  

    The method of stochastic energetics has been applied to analyze the energetic
aspects of various ratchet models (see a review \cite{ParrondoRev02,reimann} and the references therein). More recent applications are about the fluctuation theorem (FT)  \cite{Heat-maespick-JSP03,udo,ciliberto07}, the steady-state thermodynamics,  \cite{hatano-sasa}, as well as the breaking of 
fluctuation-dissipation (FD) relation \cite{FDv-harada-EPL05,FDv-harada-sasa}.
The experimental assessment of heat and work from the fluctuation of Brownian particle is also achieved \cite{THDlawUdoPRL06, ciliberto07}. The basic idea of energetics of a single stochastic trajectory has been extended to the chemical reaction\cite{ChemNetUdoJCP07} (the list is not complete).

However, along a particular trajectory of stochastic process, almost nothing has been discussed about the explicit relation between the heat in the conventional thermodynamics and the `heat' defined by the stochastic energetics. (Hereafter, we systematically use the quotation mark, `heat' etc., to mean those concept of stochastic energetics.)
  The `heat'  satisfies the first law, or the energy balance  together with the suitably defined `energy' on the {\it mesoscopic} level. The work defined there 
 satisfies the second law, that is, the positivity of the averaged irreversible work, through Jarzynski's nonequilibrium work relation \cite{Jarzynski97}. The energetics on the mesoscopic level 
 thus shows a thermodynamic structure which is proper to this level of description.
What is still missing is the link between this framework and the conventional thermodynamics, in which the heat has been measured by using a microscopic mediums (liquid/gas molecules or  conduction electrons, etc.). If the `energy' appearing in the stochastic model, such as the Langevin equation, originates simply from an external field (e.g. of optical tweezers \cite{THDlawUdoPRL06}) or from a non-entropic restoring force (e.g. of a brass-wire holding a pendulum \cite{ciliberto07}), then the `heat' can be identified with the heat in the conventional thermodynamics.
Contrastingly, the discrepancy between the `heat' and the conventional heat arises when the `energy'  contains the entropic contribution due to those microscopic degrees of freedom projected out to achieve the mesoscopic description.

To make the subject clear, let us suppose that a micron-sized magnetic bead in water
is leashed at the point $\vec{x}=0$ through a polymer chain and that 
a magnetic trap constitutes a static potential well around
$\vec{x}=\vec{a}\neq 0$. For simplicity we assume that the polymer is ideal. 
We further assume that the arrangement is such that the bead undergoes temporal
bistability, either being trapped around $\vec{x}=\vec{a}$ 
when the polymer chain is stretched to the distance
$\simeq |\vec{a}|$,  
or wandering around $\vec{x}=0$ when the chain is relaxed and fluctuating.
The main question is how much heat is released to or absorbed from the
surrounding water when the bead switches from one of the bistable states to
the other, allowing also for the change of $\vec{a}$ in time. 
The point is that, although the bistable states can be represented by
a double-well `potential' for the bead, it is only the magnetic trap 
that realizes a potential hole while the ideal chain exerts
purely entropic restoring forces. 
(We should remember that the kinetic energy of ideal chain is independent of
the chain's conformation.) 
The measured heat should depend only on the potential energy of 
the magnetic trap, which microscopic calculations should predict.
The framework of stochastic energetics predicts, however, that the
(mesoscopic) `heat' is absorbed from the environment 
whenever the bead climbs up the `potential' barrier,
and desorbed during the down-hill motion from the barrier.
Below we will show the protocol to convert the `heat' prediction
of stochastic energetics into the measurable heat, or the heat with the objectivity.
The idea is a straightforward generalization of what is known in the
equilibrium statistical mechanics. 
We will, therefore, first summarize the result of the latter discipline
(the next section
), and then go on to the stochastic dynamics
(the remaining sections
).

As for the description of ensemble behavior using the Fokker-Planck equation,
the ref. \cite{Allahverdyan1}  developed a reasoning similar to that presented in
the present paper. In particular, the authors of \cite{Allahverdyan1}  studied
two-component (fast and slow) Brownian system, investigated the
thermodynamic relation for the fast and slow components and reaches
conclusions which are consistent with those obtained in the present
paper. 

\section{Equilibrium statistical mechanics of mesoscopic variables
\label{sec:equil}}
Suppose that the total system consist of the {\it system} whose 
Hamiltonian is $H(x,y,a)$ and a heat bath of the temperature $T$.
(We could start from the whole isolated system, except that the 
argument is more complicated.)
Here $a$ stands for the external control parameter(s),
and we have divided, for the later use, the system's degrees of freedom into
two groups, $x$ and $y$.
We can define the Helmholtz free energy $F(a,\beta)$ ($\beta=(k_{\rm
  B}T)^{-1}$) through the canonical partition function, $Z(a,\beta)$
\begin{equation}
e^{-\beta   F(a,\beta)}=Z(a,\beta)={\rm Tr}_{x,y}e^{-\beta H(x,y,a)},
\label{eq:FH}
\end{equation}
where the suffices $x,y$ of ${\rm Tr}_{x,y}$ indicates the degrees of freedom
over which the trace should be taken.
We can also introduce the mesoscopic (or Landau) free energy,
$\tilde{F}(x,a,\beta)$, by 
eliminating only the degree(s) of freedom, $y$: 
\begin{equation}
e^{-\beta   \tilde{F}(x,a,\beta)}={\rm Tr}_{y}e^{-\beta H(x,y,a)}.
\label{eq:tFH}
\end{equation}
We are particularly interested in the case where 
 the variables $x$ and $y$ represent, respectively, the {\it slow} and
 {\it fast} variables of the system. 
  In the context of the example described in the Introduction, the slow variable, $x$, denotes the position of the magnetic bead, while the fast variables, $y$, describe the local movements of the monomers of the polymer chain, or even the motion of the surrounding water molecules. 
 Then $e^{-\beta \tilde{F}(x,a,\beta)}$ 
gives the relative probability density for the slow variable, $x,$ given that the parameters $a$
  and $\beta$ are fixed.

 By definition we have the relation:
\begin{equation}
e^{-\beta   F(a,\beta)}={\rm Tr}_{x}e^{-\beta   \tilde{F}(x,a,\beta)}.
\label{eq:FtF}
\end{equation}
We shall say that a quantity has the {\it objectivity}, if this quantity satisfies the following two conditions: {\it (I)} it
can be defined on the three levels
of descriptions, $\{x,y,a,\beta\}$, $\{x,a,\beta\}$ and$\{a,\beta\}$, corresponding to Eqs.~(\ref{eq:FH}), (\ref{eq:tFH}) and  (\ref{eq:FtF}), respectively, and {\it (II)} 
the magnitudes of the quantity for these descriptions are essentially the same, except for the fluctuations inherent to the description levels.
The first example is the force conjugate to the parameter $a$:
Differentiating each terms of (\ref{eq:FH}), (\ref{eq:tFH}) and 
 (\ref{eq:FtF}) with respect to $a$, we have
\begin{eqnarray}
\tilde{f}(x,a,\beta) &=&  
{\rm Tr}_{y}[e^{\beta(F-H)}\hat{f} ] \cr
f(a,\beta) = {\rm Tr}_{x} [e^{\beta(F-\tilde{F})}\tilde{f}]
&=& {\rm Tr}_{x,y}[ e^{\beta(F-H)}\hat{f}],
\label{eq:eqf} 
\end{eqnarray}
where the external force conjugated to the parameter $a$ is defined 
on the different levels,
$f(a,\beta)\equiv \partial F(a,\beta)/\partial a,$
$\tilde{f}(x,a,\beta)\equiv \partial \tilde{F}(x,a,\beta)/\partial
a$ and 
$\hat{f}(x,y,a)\equiv \partial H(x,y,a)/\partial a.$ 
The second quantity with the objectivity is the energy (not the `energy'):
Differentiating each term of (\ref{eq:FH}), (\ref{eq:tFH}) and 
 (\ref{eq:FtF}) with respect to $\beta$, we have
\begin{eqnarray}
\tilde{E}(x,a,\beta) &=&  {\rm Tr}_{y}[e^{\beta(F-H)} H] \cr
E(a,\beta) = {\rm Tr}_{x} [e^{\beta(F-\tilde{F})}\tilde{E}]
&=& {\rm Tr}_{x,y}[ e^{\beta(F-H)}H], \label{eq:eqE}
\end{eqnarray}
where $E(a,\beta)\equiv \partial [\beta F(a,\beta)]/\partial \beta$ and 
$\tilde{E}(x,a,\beta)\equiv \partial [\beta \tilde{F}(x,a,\beta)]/\partial
\beta$ stand for the energies of the system, as is $H(x,y,a)$ on the
microscopic level. 
The above relationships indicates that {(i)} it is $\tilde{F}(x,a,\beta)$
that governs the probability weight of $x$ on the mesoscopic level, while
{(ii)} it is $\tilde{E}$ whose equilibrium average over 
$x$ coincides with the thermodynamic energy $E$.
The `correction' term for the latter from the former is nothing but the 
entropic term, which we obtain by rewriting slightly the definition of 
$\tilde{E}$ mentioned above:
\begin{equation}
\tilde{E}-\tilde{F}=-T\frac{\partial \tilde{F}}{\partial T}. \label{eq:EfromF}
\end{equation}

\section{Stochastic energetics and the heat\label{sec:dyn}}
If the time-scale of the slow variable(s) $x$ is well separated from that of
fast variable(s) $y$ (as well as all the others related to the thermal
environment), and if the temperature of the environment can be regarded to be
constant, we may use the Markovian description such as  the Langevin equation to simulate the
fluctuations of $x$ near the canonical equilibrium.  
In the over-damped case, the equation writes 
\begin{eqnarray}
\gamma\frac{{\rm d}x}{{\rm d}t}=-\frac{\partial \tilde{F}(x,a,\beta)}{\partial
  x}+\xi(t),\label{eq:Leq} 
\end{eqnarray}
where $\gamma$ is the friction constant for $x$, and $\xi(t)$ is the white
 Gaussian random force with zero mean and the correlation,
$\langle \xi(t)\xi(t')\rangle=2\gamma k_{\rm B}T\delta(t-t')$ \cite{Gardiner}.
The factor $2\gamma k_{\rm B}T$ assures the canonical equilibrium distribution if $a$ is fixed.
As in the static case summarized above, it is $\tilde{F}(x,a,\beta)$ that
gives the bias for the variable $x$. 
The `energy' balance along a particular realization of the stochastic process 
writes \cite{sekimoto97}
\begin{equation}
{\rm d}\tilde{F}={\rm d}'\tilde{W}+{\rm d}'\tilde{Q},\label{eq:firstlaw}
\end{equation}
where we use ${\rm d}$ ({\it not} ${\rm d}'$) to mean the 
total differential at constant temperature, i.e., 
${\rm d}=$ ${\rm d}x\ \partial/\partial x+$ ${\rm d}a\ \partial/\partial a,$
while the work ${\rm d}'\tilde{W}$ and the `heat' ${\rm d}'\tilde{Q}$ 
brought to the system are defined by (N.B. all the
multiplications below should be interpreted as of Stratonovich type) 
\begin{equation}
{\rm d}'\tilde{W}\equiv \frac{\partial \tilde{F}}{\partial a}{\rm d}a
\end{equation}
\begin{equation}
{\rm d}'\tilde{Q}
\equiv  \left[-\gamma\frac{{\rm d}x}{{\rm d}t}+\xi(t) \right]
{\rm d}x 
=\frac{\partial \tilde{F}}{\partial x}{\rm d}x.
\end{equation}
We should remember that the eliminated degree(s) of freedom $y$ are supposed
to follow $x$ and $a$ fast enough that any non-Markov properties 
is excluded in (\ref{eq:Leq}).
It means that the the heat dissipated can be captured by the change of the 
pertinent entropy, $-\frac{\partial \tilde{F}}{\partial T}$. (A related
argument is also found in \cite{Blythe}.) 
In order to convert ${\rm d}'\tilde{Q}$ into the measurable heat, 
${\rm d}'{Q_\mathrm{m}}$,
it is, therefore, sufficient to add to both ${\rm d}'\tilde{Q}$ and ${\rm
  d}\tilde{F}$ 
the differential of the `correction' term found in (\ref{eq:EfromF}), that is 
\begin{eqnarray}
{\rm d}'\tilde{Q}&\mapsto&{\rm d}'{Q_\mathrm{m}}\equiv{\rm d}'\tilde{Q}
-T{\rm d}\frac{\partial \tilde{F}}{\partial T}   \cr
{\rm d}\tilde{F}&\mapsto&{\rm d}\tilde{E}\equiv{\rm d}\tilde{F}
-T{\rm d}\frac{\partial \tilde{F}}{\partial T}. \label{eq:main}
\end{eqnarray}
Now the `energy balance' equation (\ref{eq:firstlaw}) is converted to the 
new one that includes only the quantities with objectivity:
\begin{equation}
{\rm d}\tilde{E}={\rm d}'\tilde{W}+{\rm
  d}'{Q_\mathrm{m}}.\label{eq:firstlawbis}  
\end{equation}
This expression holds for a particular realization of the Langevin equation
(\ref{eq:Leq}), as does (\ref{eq:firstlaw}), which could be directly verified
experimentally or calculated using the original Hamiltonian $H$.
Note that the term, $-T{\rm d} (\partial \tilde{F}/\partial T),$ in (\ref{eq:firstlawbis})
is the {\it total} differential, to which both the change of $x$ and that of $a$ contribute. For cyclic processes this term has, therefore, no cumulative effects.
In the context of the fluctuation theorem about the `heat', the distribution of the (measurable) heat may deviate from that of the `heat'.

In case of the example discussed in the Introduction,
we may assign the variables $y$ to the degrees of freedoms associated to 
the monomers of the ideal chain. 
For the `potential energy', $\tilde{F}(\vec{x},\vec{a},\beta),$ 
we may write
$\tilde{F}(\vec{x},\vec{a},\beta)=$ 
$U^{\mathrm{(m)}}(\vec{x}-\vec{a})-$ $T
S^{\mathrm{(p)}}(\vec{x})$, 
where $U^{\mathrm{(m)}}(\vec{x}-\vec{a})$ represents the potential energy 
due to the magnetic trap, and $S^{\mathrm{(p)}}(\vec{x})$ is the entropy due to
the ideal polymer chain.
By substituting this form into (\ref{eq:firstlawbis}), we find the concrete
expression, term by term ($\nabla U$ denotes the gradient of $U$),
\begin{equation}
{\rm d}U^{\mathrm{(m)}}=[-{\nabla U^{\mathrm{(m)}}}(\vec{x}-\vec{a}){\rm d}\vec{a}] 
+[{\nabla U^{\mathrm{(m)}}}(\vec{x}-\vec{a}){\rm d}\vec{x}] 
\end{equation}
as it should be from the argument in the Introduction on one hand, and also as a mathematical identity on the other hand.
Experimentally, we should take account of the heat exchange with the magnetic
bead as well as the effect of polymer conformations on the solvent. 

The change of $\tilde{F}(x,a,\beta)$ through the change of $x$ is supposed to be a quasi-static work for the fast degrees of freedom, $y$. The chain should, therefore, release the heat even when the chain is spontaneously stretched near $\vec{x}=0.$ 
 This statement does not contradicts with the above analysis; it is the thermal environment that does the work to displace the bead, gathering the energy nearby. The heat $-T{\rm d} S^{(\rm p)}$ is, therefore, compensated around the system. However, if one can measure the heat 
 even closer, some local transfer of heat around the chain and the bead should be observed. 
In general, where to measure the heat ${\rm d}'Q_\mathrm{m}$ depends on to what extent we have included the fast degrees of freedom as $y$. 

\section{Case of discrete states\label{sec:dyndisc}}
It is straightforward to generalize the above analysis to the case where the system's state is discretized.  
Suppose that the probability $P_j(t)$ for the system to be in the $j$-th state obeys the 
master equation,
\begin{equation}
\frac{{\rm d} P_j}{{\rm d}t}=\sum_j \left[ P_j w_{j\to i}(a,\beta)-P_i w_{i\to j}(a,\beta)\right],
\end{equation}
where the transition rate $w_{i\to j}(a,\beta)$ from the $i$-the state to the $j$-th one writes
\cite{Lebowitz55,Lebowitz1957,Spohn,Hill89} 
\begin{equation}
w_{i\to j}(a,\beta)=\nu_0 e^{-\beta \left[\tilde{\Delta}_{i,j}(a,\beta)-\tilde{F}_i(a,\beta)\right]},
\end{equation}
where the constant $\nu_0$ is an attempting frequency, and $\tilde{\Delta}_{i,j}(a,\beta) =\tilde{\Delta}_{j,i}(a,\beta)$ is the height of the free-energy barrier between the states $i$ and $j$. 
The above form of transition rate assures the canonical equilibrium probability, $P^{\rm (eq)}_i(a,\beta)=e^{\beta (F(a,\beta)-\tilde{F}_i(a,\beta))},$ as the detailed-balance state.

The energetics of a particular trajectory corresponding to the above master equation has long been presented (see, for example, \cite{Davidson}):
If a trajectory includes the transition from the state $i_\alpha$ to the state $i_{\alpha+1}$ at the time $t_\alpha$ with $1\le \alpha \le n$ and $0<t_1<\cdots <t_n <t$, the `energy' balance 
between $t=0(\equiv t_0)$ and $t=t(\equiv t_{n+1})$ writes as follows:
\begin{equation}
\Delta \tilde{F}=\Delta' W+\Delta' \tilde{Q},
\end{equation}
with 
\begin{equation}
\Delta \tilde{F}=\tilde{F}_{i_{n+1}}(a(t),\beta)-\tilde{F}_{i_1}(a(0),\beta),
\end{equation}
\begin{equation}
\Delta' W=\sum_{\alpha=1}^{n+1}\left[
\tilde{F}_{i_{\alpha}}(a(t_\alpha),\beta)-\tilde{F}_{i_{\alpha}}(a(t_{\alpha-1}),\beta)
\right]
\end{equation}
\begin{equation}
\Delta' \tilde{Q}=\sum_{\alpha=1}^{n}\left[
\tilde{F}_{i_{\alpha+1}}(a(t_\alpha),\beta)-\tilde{F}_{i_{\alpha}}(a(t_{\alpha}),\beta).
\right]
\end{equation}
These relations correspond to (\ref{eq:firstlaw}) in the continuum case.

To transform to the balance equation with objectivity, we can again use
the correspondence relations (\ref{eq:main}):
The energy balance relation, 
\begin{equation}
{\Delta}\tilde{E}={\Delta}'\tilde{W}+{\Delta}'{Q_\mathrm{m}},\label{eq:firstlawbisdisc}  
\end{equation}
holds with  
\begin{equation}
{\Delta}\tilde{E}\equiv\Delta \tilde{F}-T\Delta \frac{\partial \tilde{F}}{\partial T}.
\end{equation}
\begin{equation}
{\Delta}'Q_\mathrm{m}\equiv\Delta' Q-T\Delta \frac{\partial \tilde{F}}{\partial T},
\end{equation}
where the {\it total} difference in the correction term is defined by 
\begin{equation}
T\Delta \frac{\partial \tilde{F}}{\partial T}\equiv 
T\left[\frac{\partial \tilde{F}_{i_{n+1}}(a(t),\beta) }{\partial T} 
-\frac{\partial \tilde{F}_{i_{1}}(a(0),\beta)}{\partial T}\right].
\end{equation}

To conclude, we have related the `heat' of the stochastic energetics with the conventional heat along a single realization of stochastic process.
For the moment, the `energy' and `heat' have only begun to be assessed experimentally \cite{THDlawUdoPRL06,ciliberto07}.
The direct measurement of the fluctuating observable heat, ${\rm d}'Q_\mathrm{m}$, will be a future experimental challenge.
The possibility to measure directly ${\rm d}'\tilde{Q}$ is an open theoretical problem.


\begin{acknowledgments}
The author thanks T. Harada and Y. Oono for valuable discussion. He also acknowledges U. Seifert for the comment on experimental situations. 
He thanks S. Sasa for the comments and for the reference \cite{Blythe},
and also a referee for drawing our attention to the ref.\cite{Allahverdyan1}.
\end{acknowledgments}

\bibliography{ken_LNP_sar}

\end{document}